\documentclass[preprint,12pt,3p]{elsarticle}
\usepackage{amssymb}
\journal {Wave Motion}

\begin {document}


\begin {frontmatter}

\title{Wave turbulence and intermittency in directional sea states}
\author[label1]{Elmira Fadaeiazar}
\address[label1]{Department of Mechanical and Product Design Engineering, FSET, Swinburne University of Technology, 3122, Hawthorn, VIC, Australia}
\author[label5]{Alberto Alberello\corref{cor1}}
\address[label5]{Department of Infrastructure Engineering, The University of Melbourne, 3010, Parkville, VIC, Australia}
\cortext[cor1]{Corresponding author}
\ead{alberto.alberello@outlook.com}
\author[label3,label7]{Miguel Onorato}
\address[label3]{Dipartimento di Fisica Generale, Universit{\`a} di Torino, 10125, Torino, Italy}
\address[label7]{INFN, Sezione di Torino, Via Pietro Giuria 1, 10125 Torino, Italy}
\author[label1]{Justin Leontini}
\author[label4]{Federico Frascoli}
\address[label4]{Department of Mathematics, FSET, Swinburne University of Technology, 3122, Hawthorn, VIC, Australia}
\author[label6]{Takuji Waseda}
\address[label6]{Department of Ocean Technology, Policy, and Environment, Graduate School of Frontier Sciences, The University of Tokyo, Tokyo, Japan}
\author[label5]{Alessandro Toffoli}

\begin {abstract}
The evolution of surface gravity waves is driven by nonlinear interactions that trigger an energy cascade similarly to the one observed in hydrodynamic turbulence. This process, known as wave turbulence, has been found to display anomalous scaling with deviation from classical turbulent predictions due to the emergence of coherent and intermittent structures on the water surface. In realistic oceanic sea states, waves are spread over a wide range of directions, with a consequent attenuation of the nonlinear properties. A laboratory experiment in a large wave facility is presented to discuss the effect of wave directionality on wave turbulence. Results show that the occurrence of coherent and intermitted structures become less likely with the broadening of the wave directional spreading. There is no evidence, however, that intermittency completely vanishes.    

\end {abstract}

\begin {keyword}
wave motion \sep wave turbulence \sep intermittency \sep structure functions \sep ocean waves
\end {keyword}

\end {frontmatter}

\section {Introduction} \label{intro}

A continuous energy cascade from large to small scales characterises isotropic and homogeneous turbulent hydrodynamic flows \citep{kolmogorov1941dissipation,kolmogorov1941local,kolmogorov1941degeneration}. At
small scales, away from the boundaries and in the limit of infinite Reynolds numbers, turbulence can be described by the scaling properties of the structure functions $S_p$. These are determined as the moments of the distribution of longitudinal velocity increments $\delta u(\tau)=u(t+\tau)-u(t)$ over small time separations $\tau$ \citep{kolmogorov1941dissipation,kolmogorov1941local}: $S_p(\tau)=\langle|\delta u(\tau)|^{p}\rangle$, where $u$ describes the flow velocity, $t$ is the time, $p$ denotes the order of the statistical moment and $\langle \cdot \rangle$ is the ensemble average. Note that reference to the time domain is made upon the Taylor's frozen turbulence hypothesis \citep{taylor1922diffusion}. 

Assuming that turbulence is statistically self-similar, there exists a unique scaling exponent $\zeta_p$ such that $S_p(\tau) \propto \tau^{\zeta_p}$. Following Kolmogorov's four-fifth law \citep{kolmogorov1941dissipation}, the exponent scales linearly as a function of the order $p$ and more specifically $\zeta_p = p / 3$ (see \citep{frisch1995turbulence} for a complete review). Interestingly enough, as the second order structure function relates to the variance of the spectral density, it can be demonstrated that in the inertial range the velocity spectrum decays as a power law of the energy $E(\omega) \propto \omega^{-\nu}$\,, with $\omega$ the angular frequency and $\nu$ the scaling exponent. Assuming a linear scaling for $\zeta_p$,  then the velocity spectrum is characterized by a power law with $\nu = 5/3$ \citep{frisch1995turbulence}. There is experimental evidence, however, that $\zeta_p$ scales nonlinearly, with departures from the Kolmogorov's prediction becoming conspicuous at higher orders \citep{frisch1995turbulence,benzi1993extended}. Divergence from the Kolmogorov's scaling relates to the presence of intermittent bursts of intense motions, which break self-similarity at small time scales. These bursts induce strongly non-Gaussian statistics for the velocity increments, with departure from Gaussianity becoming more prominent as $\tau$ becomes smaller. This phenomenon is normally known as intermittency and it characterises many different geophysical turbulent flows \citep[see, for example,][]{falkovich2001particles,de2004time,alexandrova2007solar}. As a consequence, the spectral slope deviates, albeit weakly, from the equilibrium value of $\nu = 5/3$.

On the surface of the ocean, where water depth can be assumed to be infinite (deep water), the oscillation of the surface elevation $\eta$ exhibits weakly nonlinear properties, which are responsible for an energy flux cascading towards higher frequencies. In spectral space, this corresponds to a power law that assumes the form $E(\omega)\propto \omega^{-4}$ at equilibrium \citep{zakharov1966energy,zakharov1967energy,zakharov2012kolmogorov,onorato2002freely}. This cascading behaviour resembles the one described by the Kolmogorov-type velocity spectrum in high Reynolds number flows and it is normally referred to as (weak) wave turbulence \cite{newell2001wave,nazarenko2011wave}. Despite the differences with classical turbulence, wave turbulence also exhibits intermittent properties at small scales (see e.g. \cite{falcon2007observation,falcon2010origin}). These intermittent bursts are associated with, but not necessarily limited to, the presence of coherent structures on the water surface such as sharp-crested waves, propagating breaking waves and capillarity bursts \cite{connaughton2003dimensional,yokoyama2004statistics,choi2005anomalous} and they are triggered by intense nonlinear interactions \citep{deike2015role}. Under these circumstances, the surface elevation tends to exhibits distinct non-Gaussian properties \citep{onorato2009statistical,onorato2009jfm,waseda2009evolution}. By forcing the collision of incident and reflected wave fields in a large wave basin, \citet{deike2015role} observed a variation of the intermittent behaviour due to the emergence of directional properties in the resulting wave field. The directional scatter of wave energy is a intrinsic feature of ocean waves and produces a reduction of nonlinear forcing on the water surface elevation \citep{onorato2009statistical,onorato2009jfm,waseda2009evolution}. Likewise, the surface elevation transitions to a more stable Gaussian behaviour. 

Colliding wave fields may force additional side effects such as an increase of wave amplitude and the onset of wave breaking, which can enhance nonlinear forcing and thus compensate the negative effect of directionality. Therefore, the role of directional spreading on wave turbulence is still elusive. Based on laboratory experiments in a large directional wave basin, the role of the wave directional spreading on wave turbulence and intermittency is explicitly investigated. A brief description of the experimental set up is provided in Section~\ref{exp}. The method for calculating the structure functions in wave turbulence is detailed in Section \ref{sf}. In Section \ref{res} results are presented. First, the intermittent behaviour of a unidirectional wave field is investigated. It is shown that results are consistent with previous experimental tests in e.g. \citep{deike2015role}. Then, the dependence of intermittency on the directional properties of the wave fields is discussed. Despite a notable weakening of wave turbulence with wave directionality, experimental data indicates that intermittency does not completely vanish. Concluding remarks are presented in Section~\ref{con}.    

\section {Experimental conditions} \label{exp}

\begin{figure}
\centerline{\includegraphics[width=1\textwidth]{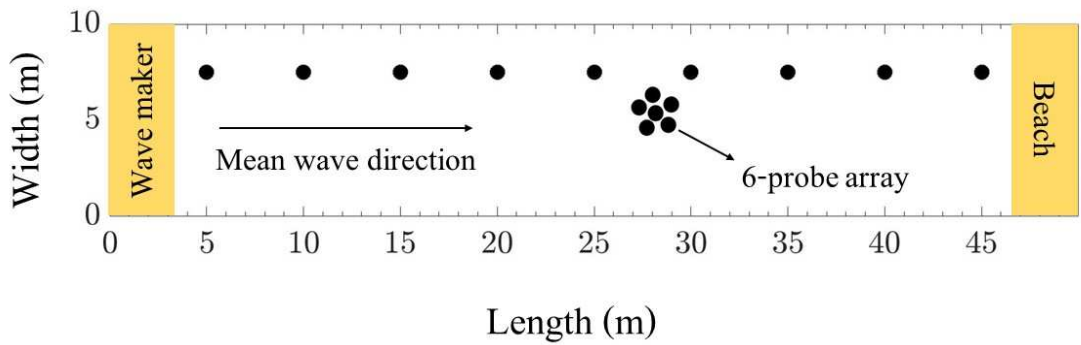}}
\caption{\label{fig:pos}Schematic of the Ocean Engineering Tank at The University of Tokyo.}
\end{figure}

Laboratory experiments have been undertaken in the Ocean Engineering tank of the Institute of Industrial Science at The University of Tokyo (Kinoshita Laboratory and Rheem Laboratory). The basin is 10 m wide, 50 m long. For the present experiment, the basin was filled with fresh water to a depth of 5\,m (see Fig.~\ref{fig:pos}). At one side, the facility is equipped with a multidirectional digitally controlled wave-maker, consisting of 32 triangular plungers, to generate random wave fields of prescribed spectral characteristics. At the opposite end, a sloping beach is mounted to absorb incoming wave energy.

The instantaneous position of the water surface was measured by nine resistance wave gauges. Probes were deployed along the tank at 5\,m intervals and at a distance of 2.5\,m from the sidewall. In order to estimate the directional properties of the wave field, one six-probe array arranged as a pentagon with one gauge at the centre of the tank was installed at 27\,m from the wave-maker (see Fig.~\ref{fig:pos}). Data were recorded at a sampling frequency is 100\,Hz (the set up for this experiment is described in details in \citep{toffoli2013excitation,toffoli2015rogue}).

The water surface was forced by the wave-maker on the basis of a pre-defined input directional wave spectrum of the type $E(\omega,\theta)=S(\omega)\cdot G(\theta)$, where $S(\omega)$ is the energy distribution in the frequency domain and $G(\theta)$ is the distribution in the directional ($\theta$) domain. The frequency spectrum was modelled according to a JONSWAP formulation \citep{komen94}:
\begin{equation}
S(\omega)=\frac{\alpha g^2}{\omega^5}\exp\left[-\frac{5}{4}\left(\frac{\omega}{\omega_P}\right)^{-4}\right]\gamma^{\exp[(\omega - \omega _P)^2/2\sigma_j^2\omega_P^2]},
\end{equation}
where $g$ is the acceleration due to gravity and $\sigma$ is a constant equal to $0.07$ for $\omega \leqslant \omega_P$ (with $\omega_P$ the angular frequency at the spectral peak) and $0.09$ for $\omega > \omega_P$; $\gamma$ is the peak enhancement factor and defines the frequency bandwidth; $\alpha$ is the Phillip's constant and it defines the energy content of the spectrum. The directional distribution $G(\theta)$ was defined with a function $G(\theta)=A_N$ $\cos^{N}(\theta)$, where $A_N$ is a normalising factor and $N$ is the directional spreading coefficient \citep{dircost714}. 

\begin{figure}
\centerline{\includegraphics[width=1\textwidth]{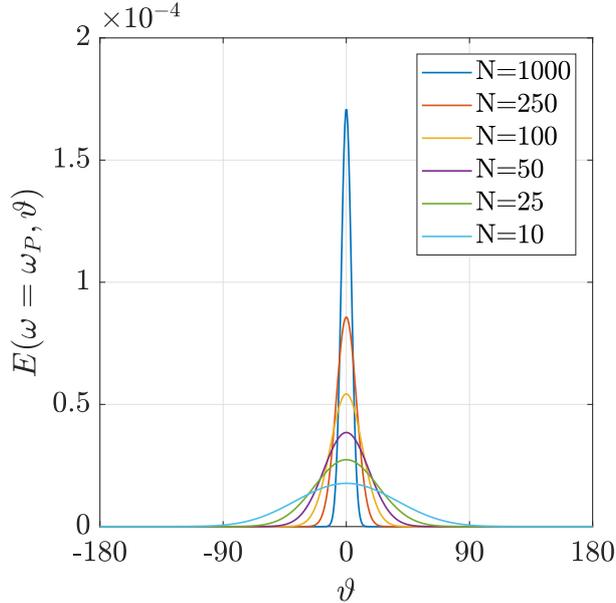}}
\caption{\label{dirspec}Analytical form of the directional distribution as a function of angle $\vartheta$.}
\end{figure}

Tests were carried out by imposing $\gamma=3$ (a typical value for ocean waves), peak wave period $T_P=0.8$\,s and $\alpha = 0.01$. This spectral configuration is characterised by a significant wave height $H_S = 0.035$\,m and a wave steepness, $\varepsilon = k_P H_S /2 = 0.11$, where $k_P$ is the wavenumber associate to the spectral peak. Note that the wave steepness is a measure of the degree of nonlinearity of the wave system. A value of 0.11, which is typical for storm conditions \citep{toffoli2005towards}, denotes a fairly nonlinear sea state (cf. \cite{onorato2009jfm}). To trace the effect of directionality on wave turbulence, different experimental tests were carried out by changing the directional spreading. Several values of the coefficient $N$ were applied, ranging from $N=1000$, a representation of an almost unidirectional sea state, to $N=10$, a broad directional distribution typical for a realistic oceanic wave field. A schematic of the different directional distributions at the spectral peak that were applied in this experiment are presented in Fig. \ref{dirspec}. It is worth noting that the selected spectral configurations are prone to nonlinear effects and the statistical properties of the surface elevation are clearly non-Gaussian if the directional spreading is narrow ($N=1000$, i.e. the wave energy remains confined around the spectral peak). A directional spreading of energy weakens nonlinearity and forces the surface elevation to transition from a strongly to a weakly non-Gaussian process (e.g. \cite{onorato2009statistical,onorato2009jfm,waseda2009evolution,toffolijfm10}).

\begin{figure}
\centerline{\includegraphics[width=1.5\textwidth]{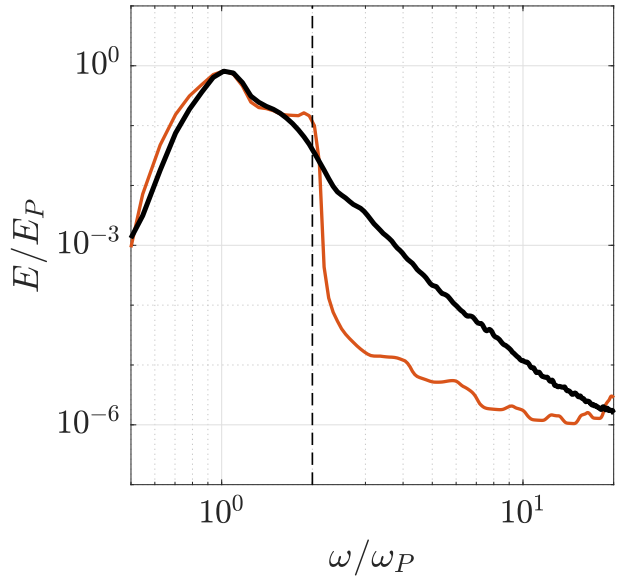}}
\caption{\label{fig:wm}Wave energy spectrum at the probe closest to the wave-maker (i.e. 5\,m from the wave-maker) in black and corresponding input voltage spectrum imposed at the wave-maker in orange. The dashed vertical line corresponds to the maximum frequency generated by the wavemaker, i.e. 2.5\,Hz.}
\end{figure}

To generate surface waves, the input spectrum was discretised into 1024 equal energy bins then converted into plunger motion superposing the Fourier modes with uniformly distributed random phases and amplitudes in the interval $[0,2\pi)$ for frequencies up to 2.5\,Hz; this limit is a mechanical constraint of the plungers. In fact, the frequency spectrum of the input signal (Fig.~\ref{fig:wm}) displays a clear drop-off for frequencies higher than 2 times the peak frequency. Nevertheless, it is worth noting that the wave spectrum already exhibits a well defined tail at 5\,m form the wave maker. For directional distribution, the Single Summation method was employed, assigning a single direction for each frequency component selecting an angle randomly from the directional spreading function as a probability density function. For each spectral configuration, a typical tests ran for 60 minutes. In order to gather enough data for a statistical analysis, two realisations for the input spectral condition were carried out with different random amplitudes and phases.

\section {Structure functions}\label{sf}

\begin{figure}
\centerline{\includegraphics[width=1.5\textwidth]{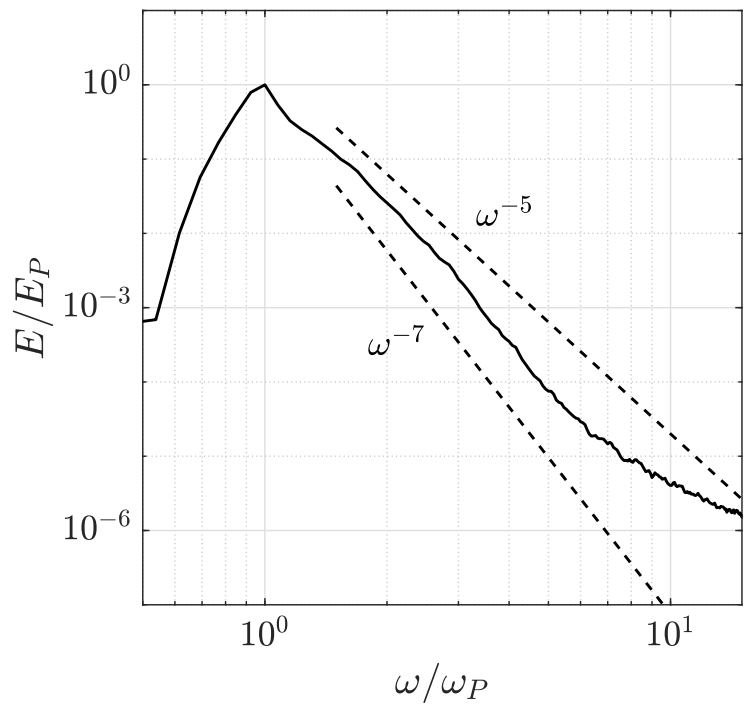}}
\caption{\label{fig:spec}Wave energy spectrum at the centre of the basin for the unidirectional wave train. The dashed lines denote the reference slope $\omega^{-5}$ and $\omega^{-7}$ respectively.}
\end{figure}

The theoretical framework developed for classical hydrodynamic turbulence \citep{frisch1995turbulence} is commonly adapted to wave turbulence, but some notable differences stand. First and foremost, the spectrum of the surface elevation develops a spectral tail $\omega^{-4}$ due to the emergence of four wave interactions driven by the weakly nonlinear properties of water waves \citep{zakharov1966energy,zakharov1967energy,zakharov2012kolmogorov,onorato2002freely,toffoli2017wind}. We recall that the equilibrium slope of the spectrum relates to the value assumed by the second order structure function, which is proportional to the variance of the surface elevation increments in wave turbulence \citep{falcon2010revealing}. The surface elevation increments have to be computed at the second order to restore the stationarity of the statistical properties of the surface elevation increments and, thus, apply the classical turbulence theory for such a spectral slope \citep{falcon2010origin,deike2015role,falcon2010revealing}. However, the equilibrium tail can shift toward $\omega^{-5}$ when wave breaking dissipation becomes dominant. In laboratory experiments, moreover, mechanically generated waves often exhibit a spectral tail with slope $\omega^{-5}$ or even steeper \citep{deike2015role,onorato2009statistical,waseda2009interplay,aubourg2017three}. Wave spectra for the current experiments display in fact a spectral tail between $\omega^{-5}$ and $\omega^{-7}$ (see also similar experiments in the same facility in \citep[][]{waseda2009evolution,waseda2009interplay}). As an example, the spectrum calculated for the unidirectional wave train is shown in Fig.~\ref{fig:spec}. A similar behaviour was observed for directional wave fields. We remark that the spectral tail is not forced by the wave-maker, but naturally emerges as a result of wave-wave interaction that triggers the wave turbulence cascade (see Fig.~\ref{fig:wm}). The imposed discrete spectral energy naturally spreads in directions as well, resulting in a formation of the continuous directional wave field.
Under the present experimental conditions, third-order differences of the surface elevation must be used \citep{falcon2010revealing}. These are defined as
\begin{equation}
\delta^{(3)}_\tau\eta=\eta(t+3\tau)-3\eta(t+2\tau)+3\eta(t+\tau)-\eta(t),
\end{equation}
where $\eta$ denotes the surface elevation. The corresponding structure functions can thus be expressed as
\begin{equation}
S_p(\tau)=\langle|\delta_\tau^{(3)} \eta(\tau)|^{p}\rangle.
\end{equation}

In classical turbulence and wave turbulence alike, determination of the scaling exponents from the increments is challenging due to uncertainties related to viscous effects (finiteness of the Reynolds number), inhomogeneity, anisotropy, violation of Taylor's hypothesis, and experimental errors \citep{frisch1995turbulence,anselmet1984high}. Whereas low order structure functions ($p\leq 4$) are less prone to uncertainties, errors become significant at high orders. To evaluate the scaling exponent $\zeta_p$ more accurately, the Extended Self-Similarity (ESS) hypothesis can be employed \citep{benzi1993extended}. In classical turbulence, this hypothesis states that, given the exact exponent $\zeta_3=1$ \citep{kolmogorov1941dissipation,kolmogorov1941local}, the other ones can be derived relatively to $S_3$, i.e. $S_p(\tau)\propto S_3(\tau)^{\zeta_p}$. Validity of the relative scaling (relative to $p=3$) extends beyond the inertial range allowing for a more robust estimation of $\zeta_p$ \citep{benzi1993extended}. The exact exponent, $\zeta_3=1$, does not hold in wave turbulence and the second order structure functions has been used as a reference to apply the extended self similarity (i.e. $S_p \propto S_2^{\zeta_p}$) \citep{chibbaro2017weak} and overcome experimental uncertainties. Therefore, $\zeta_p$ denotes a relative scaling exponent in this case. By using $S_2$ as a reference structure function, the scaling exponent becomes $\zeta_p = p/2$ in absence of intermittent bursts.

\section {Results}\label{res}

Without loss of generality, a detailed data analysis of wave turbulence and intermittency is presented for unidirectional waves only. In order to include a sufficiently large data set and statistical significance, records from the six-probe array at the centre of the tank are considered herein.

\begin{figure}
\centerline{\includegraphics[width=1.2\textwidth]{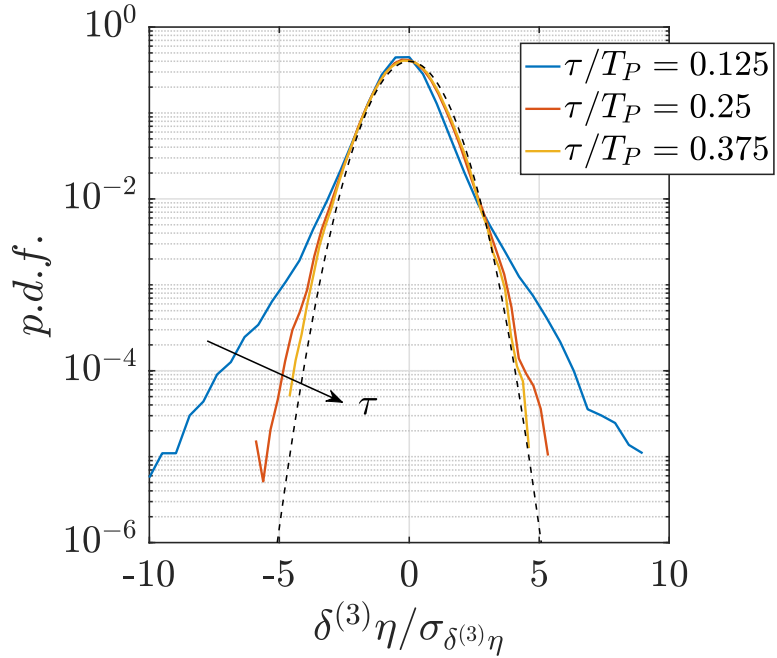}}
\caption{\label{fig:pdf}Example of the probability density function of the third order surface elevation increments for three different time lags $\tau$.}
\end{figure}

A first indication of the properties of wave turbulence can be obtained by the probability density function of the third order surface increments (Fig.~\ref{fig:pdf}). For large separation distances ($\tau$), the increments show a quasi-Gaussian behaviour. For small separations, on the other hand, the probability density function displays heavy tail statistics, which is symptomatic of an intermittent behaviour \citep{falcon2010revealing}. Statistics of the increments reveal that intermittency is localised at the smallest scales (i.e. it is still clearly noticeable at $\tau/T_P=1/8$), while it disappears almost completely for separation larger than a quarter of the wave period.

\begin{figure}
\centerline{\includegraphics[width=1\textwidth]{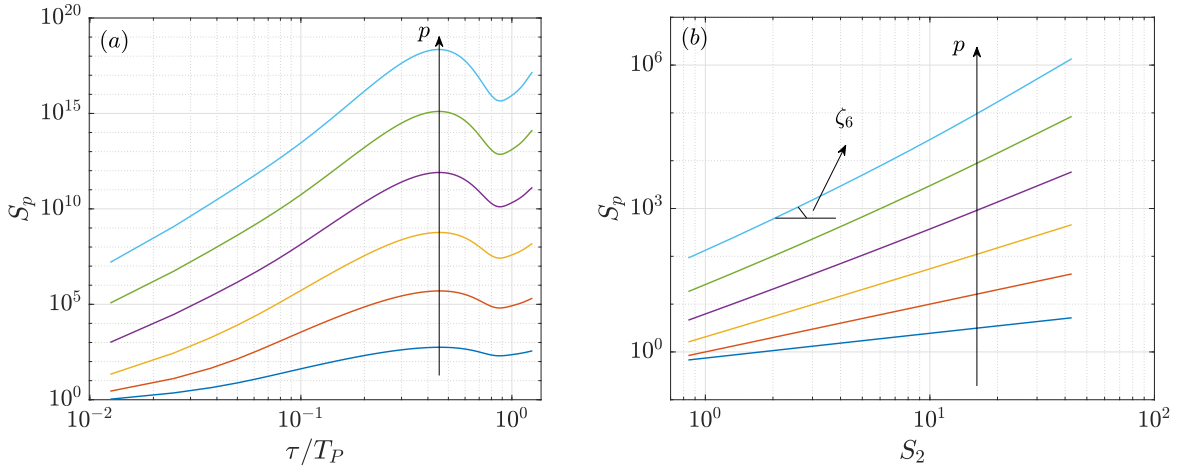}}
\caption{\label{fig:stress}Example of the structure function of the third order surface elevation increments up to the order $p=6$ as a function of the dimensionless time lag ($a$) and as a function of the second order structure function ($b$).}
\end{figure}

The structure functions $S_p$ of the surface elevation increments are estimated for orders up to $p = 6$. An example of $S_p$ as a function of the normalised time scale $\tau / T_P$ is shown in Fig.~\ref{fig:stress}a. In a double logarithmic plane, the structure functions increase with the time separation up to $\tau/T_P \approx 0.5$. The subsequent drop in the structure functions at large $\tau$ can be attributed to the periodic nature of water waves. Measurements separated by one dominant wave period are likely to show high correlation. The structure functions only display a linear trend over a very limited range of time lags, making the computation of the exponent $\zeta_p$ uncertain especially for high orders $p$. The relative scaling exponent derived from the extended self similarity approach is used (see Fig.~\ref{fig:stress}b).

\begin{figure}
\centerline{\includegraphics[width=1.2\textwidth]{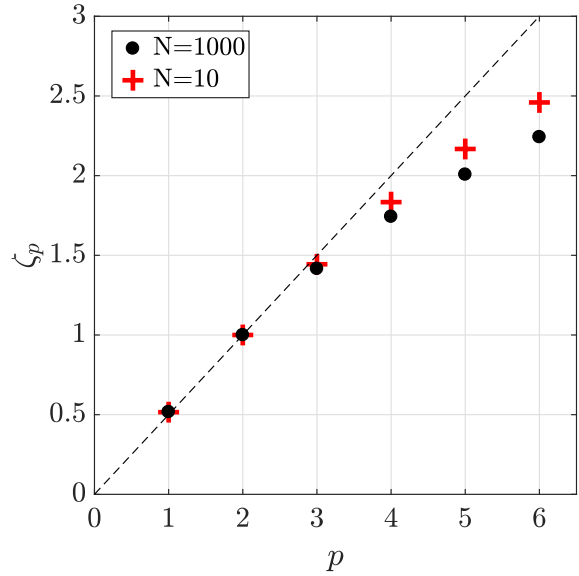}}
\caption{\label{fig:strf}Exponent of the structure function as a function of the order $p$ as calculated in the middle of the basin for unidirectional waves and directionally spread waves. The dashed line denotes the line $\zeta_p=p/2$, corresponding to a no intermittent behaviour.}
\end{figure}

A least square method fit is applied to the data points in the plane $S_2$ vs. $S_p$ (Fig.~\ref{fig:stress}b) to extract the slope of the curves. This directly provides the relative scaling exponent $\zeta_p$. Exponents up to the order $p=6$ are shown in Fig.~\ref{fig:strf}. Deviation from the predicted $\zeta_p=p/2$ for high order $p$ confirms departure from classical turbulence predictions and the emergence of intermittent behaviour, which become more evident with the increase of the order $p$. Results for unidirectional waves are consistent with observations reported in \citet{deike2015role}.

A detailed analysis of the structure function for directional sea states indicates the persistence of intermittency despite the broadening of the spectral shape. An example of the exponents $\zeta_p$ as a function of the order $p$ for the broadest directional sea  ($N=10$) is reported in Fig.~\ref{fig:strf}. 

\begin{figure}
\centerline{\includegraphics[width=1.0\textwidth]{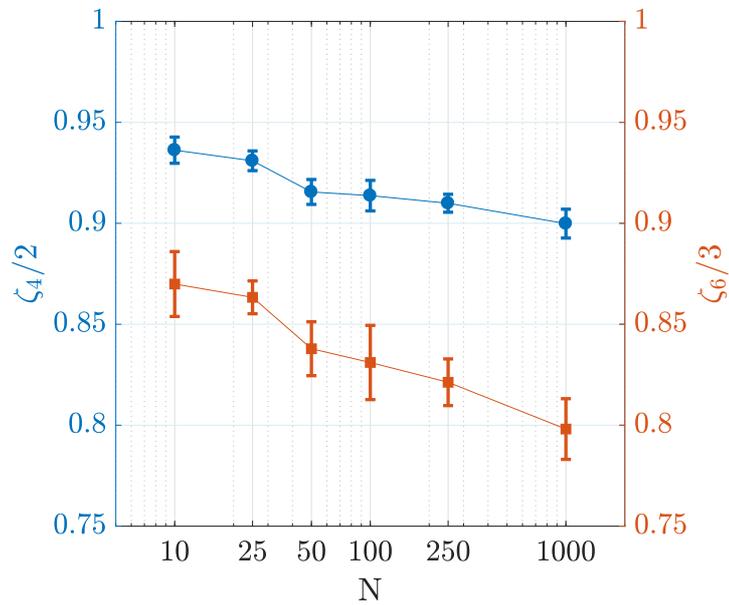}}
\caption{\label{fig:str46}Structure function exponent as a function of the directional spreading parameter $N$. On the left axis, teal circles denotes $\zeta_4$ normalised by 2 (i.e. reference value for $p=4$ in absence of intermittency). On the right axis, orange squares denotes $\zeta_6$ normalised by 3 (i.e. reference value for $p=6$ in absence of intermittency). The error bar is computed as two times the standard deviation computed for the six probes in the array.}
\end{figure}

The role of the wave directional spreading on the strength of intermittency is summarised in Fig.~\ref{fig:str46}, where even exponents $\zeta_4$ and $\zeta_6$ are presented against the spreading coefficient $N$. The respective reference values of 2 and 3 in the absence of intermittency are used as normalising factors. The emergence of intermittency in the unidirectional sea state ($N=1000$), in this respect, is denoted by a notable deviation from the benchmark values of 1 (no intermittency) and specifically $\zeta_4/2 = 0.9$ and $\zeta_6/3 = 0.8$. Departures from benchmark, however, gradually lessens as the wave directionality broadens. For $N$ approaching 10, $\zeta_4/2 = 0.94$ and $\zeta_6/3 = 0.87$. The trend is more apparent for higher exponents, as expected.

The dependence of intermittency strength on directional wave properties mirrors the one observed for the statistical properties of the surface elevation. Waves transition from a strongly to a weak non-Gaussian state when the wave field spreads on directions \cite{onorato2009statistical,onorato2009jfm,waseda2009evolution,toffolijfm10}. Nevertheless, even for the most directional wave field ($N=10$), the normalised $\zeta_4$ and $\zeta_6$ remain smaller than the benchmark value of 1. This indicates clearly that the intermittency still persists despite directionality, albeit weak.
 
\section {Conclusions} \label{con}

Records of the water surface elevation in a directional wave tank are used to explore the effect of different spectral wave conditions on the properties of wave turbulence and intermittency. Random directional sea states were generated mechanically (i.e. with the movement of a plunger) by imposing a desired input directional wave spectrum at the wave-maker. After generation, the wave field propagated freely along the basin. Nonlinear wave interactions were the only driving factor on wave dynamics.

Results indicates that the surface elevation displays an intermittent behaviour with deviation from classical wave turbulence predictions. Nevertheless, experiments demonstrate that intermittency strength reduces with the increase of the level of wave directionality. This behaviour is consistent with the fact that the available energy for each spectral component is reduced. In this regard, it is worth remarking the analogy between statistical properties of the surface elevation and intermittency strength. However, although the surface elevation transitions from a strongly to a quasi-Gaussian process, intermittency still persists without completely vanishing.

\section*{Acknowledgements} 
The experiments were supported by the JSPS Fellowship for Research in Japan Program, Grants-in-Aid for Scientific Research of the JSPS and the International Science Linkages (ISL) Program of the Australian Academy of Science. M.O. received support from Progetto di Ricerca d'Ateneo CSTO160004 and the ``Departments of Excellence 2018--2022'' Grant awarded by the Italian Ministry of Education, University and Research (MIUR, L.232/2016). A.A. and A.T. acknowledge support from the Air-Sea-Ice Lab Project. M.O. is grateful Dr. B. Giulinico for discussions.


\end {document}